\documentclass[a4paper]{jpconf}
\usepackage{graphicx}
\usepackage{url}
\usepackage{subfig}

\begin{document}
\title{XSL: The X-Shooter Spectral Library}

\author{Yanping Chen$^{1}$, Scott Trager$^{1}$, Reynier Peletier$^{1}$
  and Ariane Lan\c{c}on$^{2}$}

\address{$^{1}$Kapteyn Astronomical Institute, University of
  Groningen, Postbus 800, 9700 AV, Groningen, Netherlands}
\address{$^{2}$Observatoire Astronomique, 11, rue de l'Universit\'e,
  F-67000 Strasbourg, France}

\ead{yanping@astro.rug.nl}

\begin{abstract}
We are building a new spectral library with the X-Shooter instrument
on ESO's VLT: XSL, the X-Shooter Spectral Library.  We present our
progress in building XSL, which covers the wavelength range from the
near-UV to the near-IR with a resolution of $R\sim10000$. At the time
of writing we have collected spectra for nearly 240 stars.  An
important feature of XSL is that we have already collected spectra of
more than 100 Asymptotic Giant Branch stars in the Galaxy and the
Magellanic Clouds.

\end{abstract}

\section{Introduction}

Stellar population models are powerful tools which are widely used to
study galaxy evolution.  Using these models, one can determine galaxy
ages, metallicities and abundances.  Spectral libraries are an
integral component of stellar population models (e.g. \cite{BC03}).  A
spectral library gives the behavior of individual stellar spectra as
function of effective temperature ($T _{\mathrm{eff}}$), gravity
($\log g$) and metallicity ([Fe/H]). A stellar population model
integrates these spectra together with a set of stellar isochrones and
an initial mass function to produce a model spectrum of an entire
population.  In order to reproduce galaxy spectra as precisely as
possible, one requires a comprehensive stellar spectral library that
covers the entire desired parameter space of $T _{\mathrm{eff}}$,
$\log g$, and [Fe/H].  Moreover, extended wavelength coverage is
strongly desirable, because different stellar phases contribute their
light in different bands. For instance, bright giants contribute more
light in the near-infrared than the faint giant and subgiant stars,
while in the optical, the situation is reversed \cite{Frogel88}.
These Asymptotic Giant Branch (AGB) stars dominate the light of
intermediate-aged and old stellar populations in the near-infrared but
are unimportant in the optical (e.g., \cite{Maraston05,Conroy10}).
Detecting their presence requires \emph{broad wavelength coverage} in
both the target and model spectra.

Stellar spectral libraries can be classified into empirical and
theoretical libraries, depending on how the library is obtained. Both
theoretical and empirical libraries have improved in recent years. The
most widely used theoretical libraries in stellar population models
are those of \cite{Kurucz93}, \cite{Munari05}, \cite{Gustafsson08},
\cite{Coelho05}, \cite{Coelho07}, and \cite{Martins05}. Theoretical
libraries have the advantage of (nearly) unlimited resolution and
selectable abundance patterns -- not only scaled-solar abundances but
also non-solar patterns.  Unfortunately, theoretical libraries suffer
from systematic uncertainties, as they rely on model atmospheres and
require a reliable list of atomic and molecular line opacities
\cite{Coelho05}.  Empirical stellar libraries, on the other hand, have
the advantage of being drawn from real, observed stars and therefore
do not suffer this limitation; however they frequently have relatively
low resolution (with a few exceptions; see below) and are unable to
reproduce the indices measured in giant elliptical galaxies
\cite{Reynier89, Worthey92}, because they are based on local stars with
typical Milky Way disk abundance patterns.

In this proceeding we present the X-Shooter Stellar Library, XSL.  XSL
is being obtained using the new X-Shooter three-arm spectrograph on
ESO's VLT \cite{Vernet10}.  XSL has the unique advantage of simultaneously
acquiring spectra covering the near-ultraviolet up to the near-infrared.
Furthermore, the fact that X-Shooter is mounted on an 8.2-m telescope
allows us to include faint objects in stars in the Galactic bulge and
the Magellanic Clouds for the first time, along with stars in the
Galactic disk and halo, at moderate spectral resolution ($R\sim10000$).

\subsection{Previous stellar libraries}

\begin{table}
\caption{\label{tablib}Some previous stellar libraries}

\begin{center}
\lineup
\begin{tabular}{*{5}{l}}
\br
Library&Resolution&Spectral range&Number &Reference\cr
&R=$\lambda/\Delta\lambda$&(nm)&of stars&\cr
\mr
STELIB & 2000 & 320-930 & 249 &\cite{LeBorgne03}\cr
ELODIE & 10000 & 390-680 & 1300 &\cite{Prugniel01,Prugniel04,Prugniel07}\cr
INDO-US & 5000 & 346-946 & 1237&\cite{Valdes04}\cr
MILES & 2000 & 352-750 & 985 &\cite{miles06}\cr
IRTF-SpeX & 2000 & 800-2500& 210 &\cite{Rayner09}\cr
NGSL & 1000 & 167-1025& 374 &\cite{Gregg06}\cr
UVES-POP & 80000 & 307-1030& 300 &\cite{Bagnuolo03}\cr
LW2000 & 1100 & 500-2500& 100 &\cite{LW2000}\cr
\br
\end{tabular}
\end{center}
\end{table}

We begin with a review of several previous empirical stellar libraries
and their principal features, listed in Table~\ref{tablib}.  In the
optical, we have (among others) Lick/IDS \cite{Worthey94}, MILES
\cite{miles06}, ELODIE \cite{Prugniel01,Prugniel04,Prugniel07}, STELIB
\cite{LeBorgne03}, NGSL \cite{Gregg06}, and the Pickles library
\cite{Pickles85}.  Libraries in the near-IR are a challenging task,
but pioneering work has been done by Lan\c{c}on and Wood \cite{LW2000}
(LW2000), Rayner et al.\ \cite{irtf09} (IRTF-SpeX), and
M\'{a}rmol-Queralt\'{o} et al.\ \cite{Marmol-Queralto}. However,
extended-wavelength-coverage spectral libraries at moderate resolution
are still largely missing.

STELIB \cite{LeBorgne03} has been used to construct the widely used
stellar population synthesis models of Bruzual \& Charlot 2003
(\cite{BC03}, here after BC03).  This library has a wide wavelength
range, 3200--9300 \AA, at modest resolution ($R\sim2000$).  The
spectra of local galaxies have been reproduced by the models of BC03
through this library \cite{Gallazzi05}. However, stars with low
and high metallicites are sparse in this library, leading to problems
in some regimes, like modeling old, metal-poor globular cluster
spectra \cite{Koleva08}.

The ELODIE library \cite{Prugniel01,Prugniel04,Prugniel07} is a
moderate-resolution spectral library with $R\sim10000$. It has been
applied in the P\'{E}GASE-HR synthetic model \cite{LeBorgne04}. The
atmosphere parameter coverage of this library had been improved
through its updated version, but the wavelength range is still limited
in the optical region (4100--6800 \AA).

The INDO-US library \cite{Valdes04} includes a large number of stars
(1273), covers a fair range in atmosphere parameters with moderate-resolution
($R\sim5000$) . Atmospheric parameters of its stars are given in
Wu et al. \cite{Wuyue11}. The data have been used for automated spectral
classification of high-resolution spectra over a wide wavelength range
\cite{Pickles07}.  The problem of this library is the lack of accurate
spectrophotometry, making inclusion in stellar population models
problematic.

The MILES library \cite{miles06} is widely used in stellar population
models. This library profits from its good physical parameter
coverage, careful flux calibration, and a large number of stars (985),
enabling stellar population models to predict metal-poor or metal-rich
systems. The MILES library's wavelength range covers 3500--7500 \AA\
at modest resolution ($R\sim2000$) \cite{Falcon11}.

The Next General Spectral Library (NGSL\footnote{See
  \url{http://archive.stsci.edu/prepds/stisngsl/}}, \cite{Gregg06}) is
a low resolution ($R\sim1000$) stellar library obtained with STIS on
HST.  It contains $\sim400$ stars having a wide range in metallicity
and age. Since its spectra were observed in space, this library does
not suffer from telluric absorption or seeing variations, and hence
has excellent absolute spectrophotometry. Other features of NGSL are
its broad wavelength coverage, from the space UV to 1 $\mu$m, and high
signal-to-noise.  There is also a high-resolution ($R\sim40000$,
covering 3600--11000 \AA) extension of NGSL's southern stars taken
with the UVES spectrograph of ESO's VLT (Hanuschik et al., in prep.).

The UVES Paranal Observatory Project (UVES-POP, \cite{Bagnuolo03}) is a
library of spectra of $\sim300$ nearby bright stars taken with
UVES. The spectra cover the optical region at $R\sim80000$ and have a
typical S/N ratio is 300--500 in the $V$ band. The library has been
flux calibrated but has not been corrected for the severe telluric
absorption in the red.  For stellar population models the sample is
incomplete, since it only contains bright stars around solar
metallicity.

In the near-infrared range, the IRTF Spectral Library (IRTF-SpeX,
\cite{Rayner09}) contains 200 stars observed with the cross-dispersed
infrared SpeX spectrograph on IRTF at a resolving power of
$R\sim2000$. The largest spectral library of very cool supergiants and
giants over the wavelength range $\lambda=0.5$--2.5 $\mu$m was
published by Lan\c{c}on \& Wood (LW2000, \cite{LW2000}), which
contains $\sim100$ stars taken at a resolution of $R=1100$, including
observations over multiple phases. Unfortunately, cool giants are
variable in optical and NIR, so simultaneous observations are
required.  Libraries with more limited spectral coverage in the
near-infrared, like the 2.3 $\mu$m library of Marm\'ol-Queralt\'o et
al.\ \cite{MQ08}, also exist, but are limited to a small number of
spectral features (like the CO bandhead).

\section{XSL}

X-Shooter \cite{Vernet10} was built by a consortium of 11 institutes
in Denmark, France, Italy and the Netherlands, together with ESO. It
is currently mounted on UT2 on ESO's VLT.  A unique capability of
X-Shooter is that it collects spectra in the wavelength range from the
near-ultraviolet to the near-infrared -- 300--2500 nm -- through its
three arms \emph{simultaneously}. This character is extremely useful
for observing variable stars, especially very cool stars -- like
stars -- whose spectra vary substantially during their pulsation
cycles.

\subsection{Sample selection}

XSL targets were selected from many of the above libraries as well as
supplementary literature sources. We took stars from Lick/IDS, MILES,
and NGSL to cover $T _{\mathrm{eff}}$, $\log g$, and [Fe/H] as
uniformly as possible.  However, these libraries mostly lack the cool,
bright stars so important in the near-infrared.  For this purpose we
selected AGB and LPV stars from LW2000 and IRTF-SpeX with declination
$<35\,^{\circ}$ marked M, C or S-stars. Long-period variable (LPV)
stars were also collected from the LMC \cite{HW90} and SMC
\cite{Cioni03}. Red supergiant stars were taken from the lists of
LW2000 and \cite{Levesque05,Levesque07}. To cover metal-rich stars
with abundances similar to giant elliptical galaxies, we also included
Galactic Bulge giants from the samples of
\cite{BMB,Groenewegen05}. Our intention is that XSL will contain about
600 stars at the end of the survey.

\subsection{Observations and data reduction}

\begin{figure}
\begin{minipage}{18pc}
\includegraphics[width=18pc]{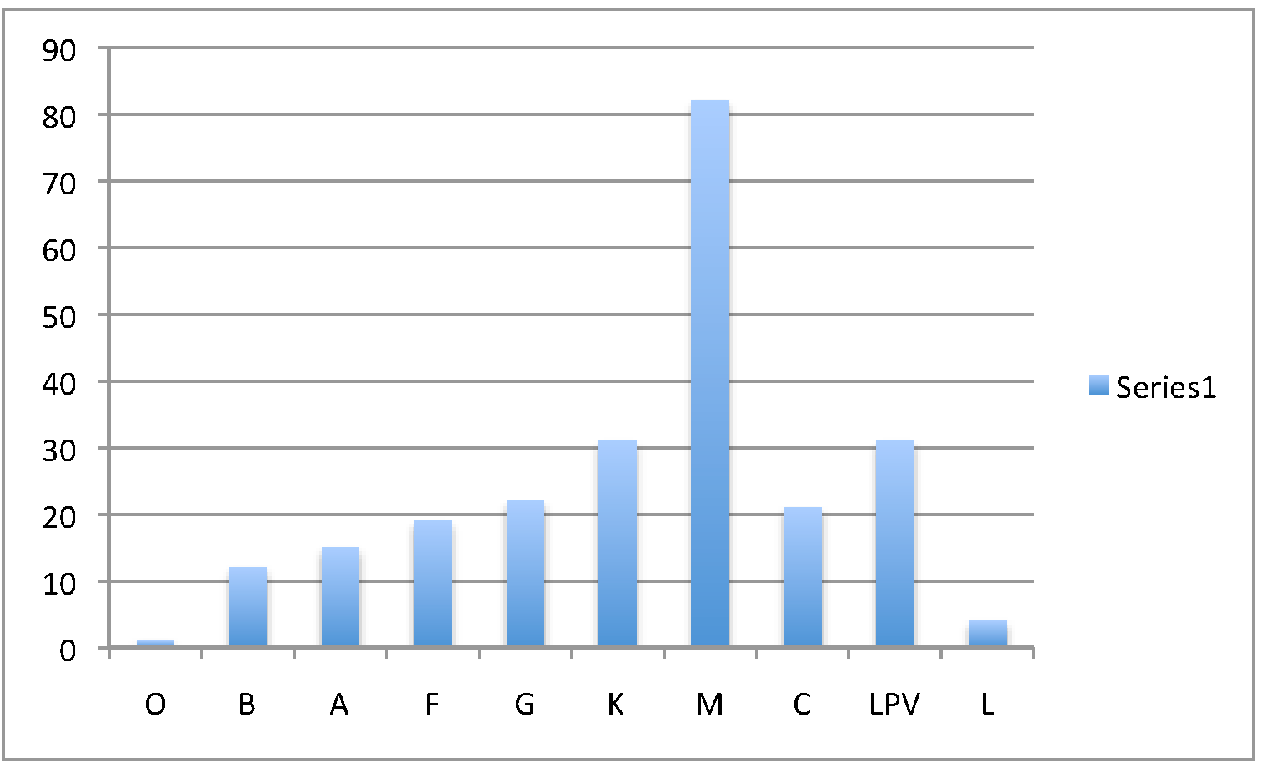}
\caption{\label{obaf}Distribution of spectral types of the main XSL
  samples (excluding telluric calibrators), retrieved from SIMBAD or
  based on educated guesses from the source library or atmospheric
  parameters.}
\end{minipage}\hspace{2pc}%
\begin{minipage}{18pc}
\includegraphics[width=18pc]{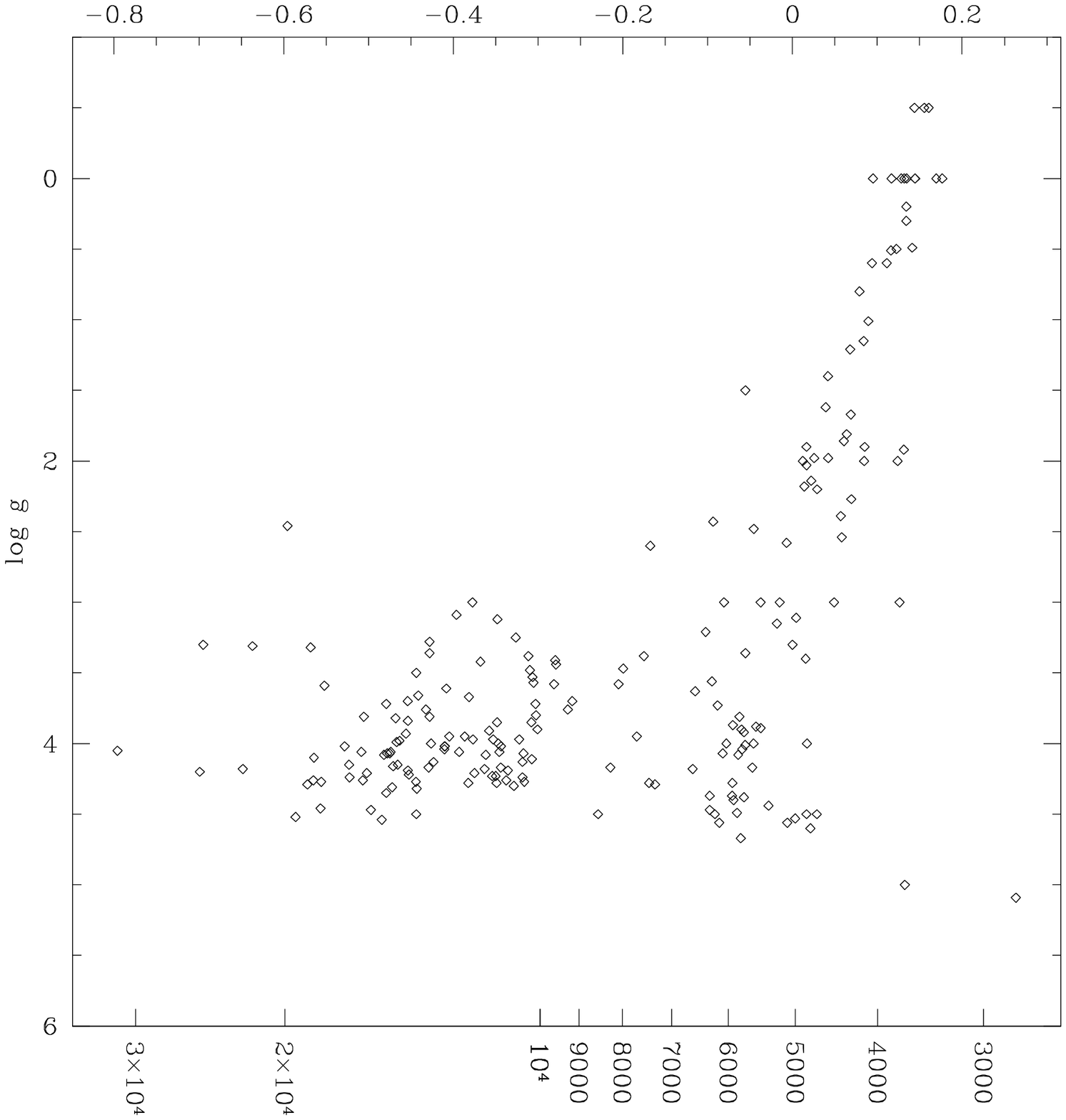}
\caption{\label{param}HR diagram of the 114 X-Shooter library stars
  with known $T_{\mathrm{eff}}$ and $\log g$. Bottom axis is $T_{\mathrm{eff}}$;
 top axis is log $\theta$, where $\theta$ = 5040/$T_{\mathrm{eff}}$ .}
\end{minipage}
\end{figure}

To date, 251 observations of 236 unique stars from the XSL input
catalog have been made. Fig.~\ref{obaf} shows the
distribution of our sample stars with different stellar types. We have
collected a large sample of M, C, and LPV stars, interesting for many
different cool star studies. In Fig.~\ref{param} we show those sample
stars with known stellar parameters in an HR diagram.

The targets were observed with X-Shooter with a
$0.5^{\prime\prime}\times11^{\prime\prime}$ slit in the UVB arm
(3000--6000 \AA, $R=9100$), a
$0.7^{\prime\prime}\times11^{\prime\prime}$ slit in the VIS arm
(6000--10200 \AA, $R=11000$), and a
$0.6^{\prime\prime}\times11^{\prime\prime}$ slit in the NIR arm
(1--2.5 $\mu$m, $R=8100$). Stars were nodded along the slit to remove
the sky background, a serious problem in the NIR arm even for very
short exposures. In addition, for every star a wide
($5.0^{\prime\prime}$) slit spectrum was taken to allow for excellent
flux calibration (see, e.g., \cite{miles06}).

\begin{figure}[h]
\includegraphics[width=28pc,angle=90]{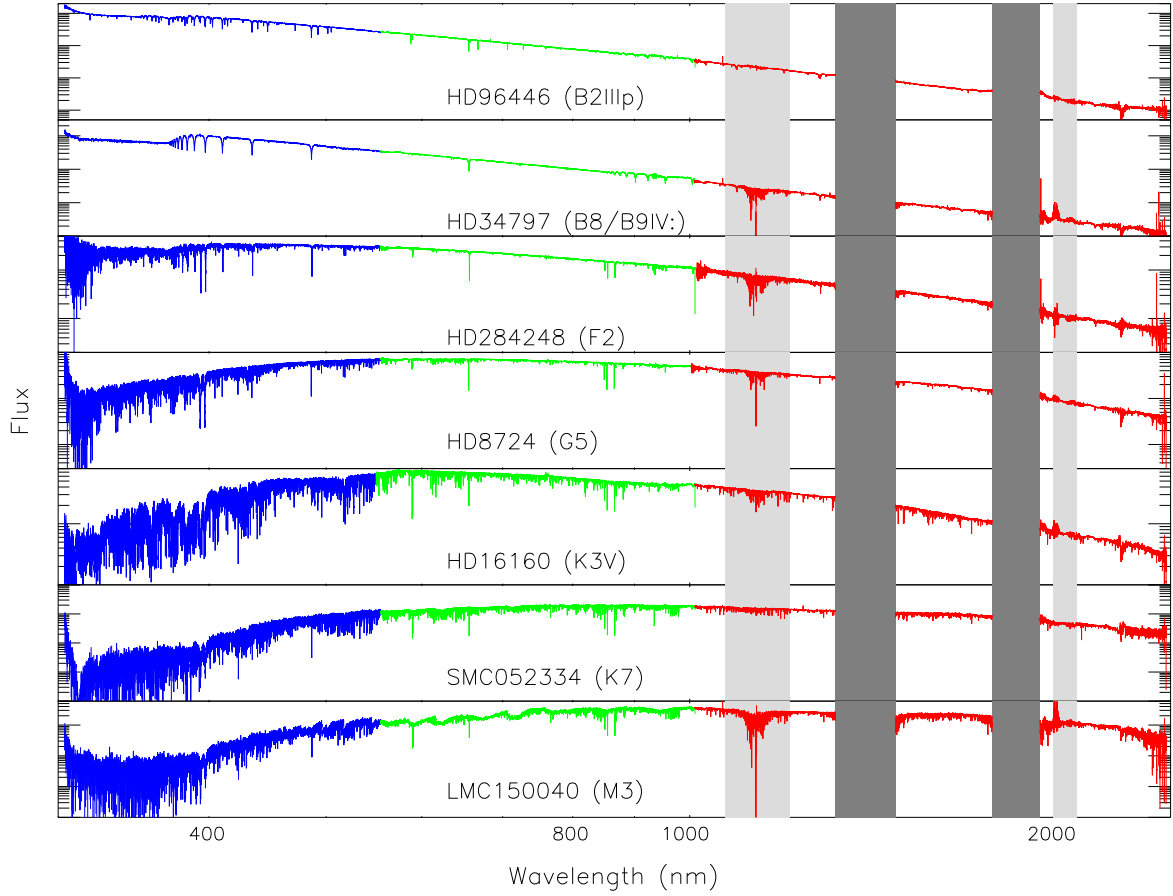}\hspace{2pc}%
\caption{\label{nstar} A sample of XSL stellar spectra sorted by temperature.
 Grey bars cover areas of severe (light grey) or nearly complete (dark grey)
 telluric absorption in the NIR arm.}

\end{figure}

The basic data reduction was performed with the public release of the
X-Shooter pipeline version 1.1.0, following its standard steps up to
the production of 2D spectra, including bias and/or dark correction,
flat-fielding, wavelength calibration, sky subtraction.
We then used IRAF's \texttt{twodspec.apextract} routine
to extract 1D spectra from the 2D spectra. We show typical X-Shooter
library spectra in Fig.~\ref{nstar}, which contains several spectral
types from B2 to M3. In this figure flux calibration has been applied to
all the three arms. The UVB arm is shown in blue, the VIS arm in green and the NIR
arm in red. Telluric corrections have been applied to the VIS and NIR arms.
Grey bars demonstrate the telluric regions in the NIR arm, in which the
light grey bars cover the areas of severe telluric absorption while the dark
grey ones cover the areas nearly complete telluric absorption.

\subsection{Telluric correction}

Since X-Shooter is a ground based instrument, correction for telluric
absorption in the VIS and NIR arm spectra is very
important. For this reason a hot star (typically a mid-B dwarf) was
observed after every science object at a similar
airmass. Unfortunately, the telluric absorption lines change strength
on timescales shorter than the ``long'' exposure times ($\geq90$ seconds)
of faint XSL stars and the total observational overhead time of
$\sim900$ seconds, resulting in a noisy telluric correction. To
optimize the telluric correction, we built a telluric stellar spectrum
library, in which the hot stars were carefully wavelength calibrated.
First we masked out strong H and other lines from the telluric region,
since we know that B and other hot stars usually have strong H lines
which must be corrected before the spectra can be used as telluric
templates.  We then normalized the masked spectra to create our
telluric standard library.






\begin{figure}
    \centering
    \subfloat[]
       {\label{eg0}\includegraphics[angle=90,scale=0.25]{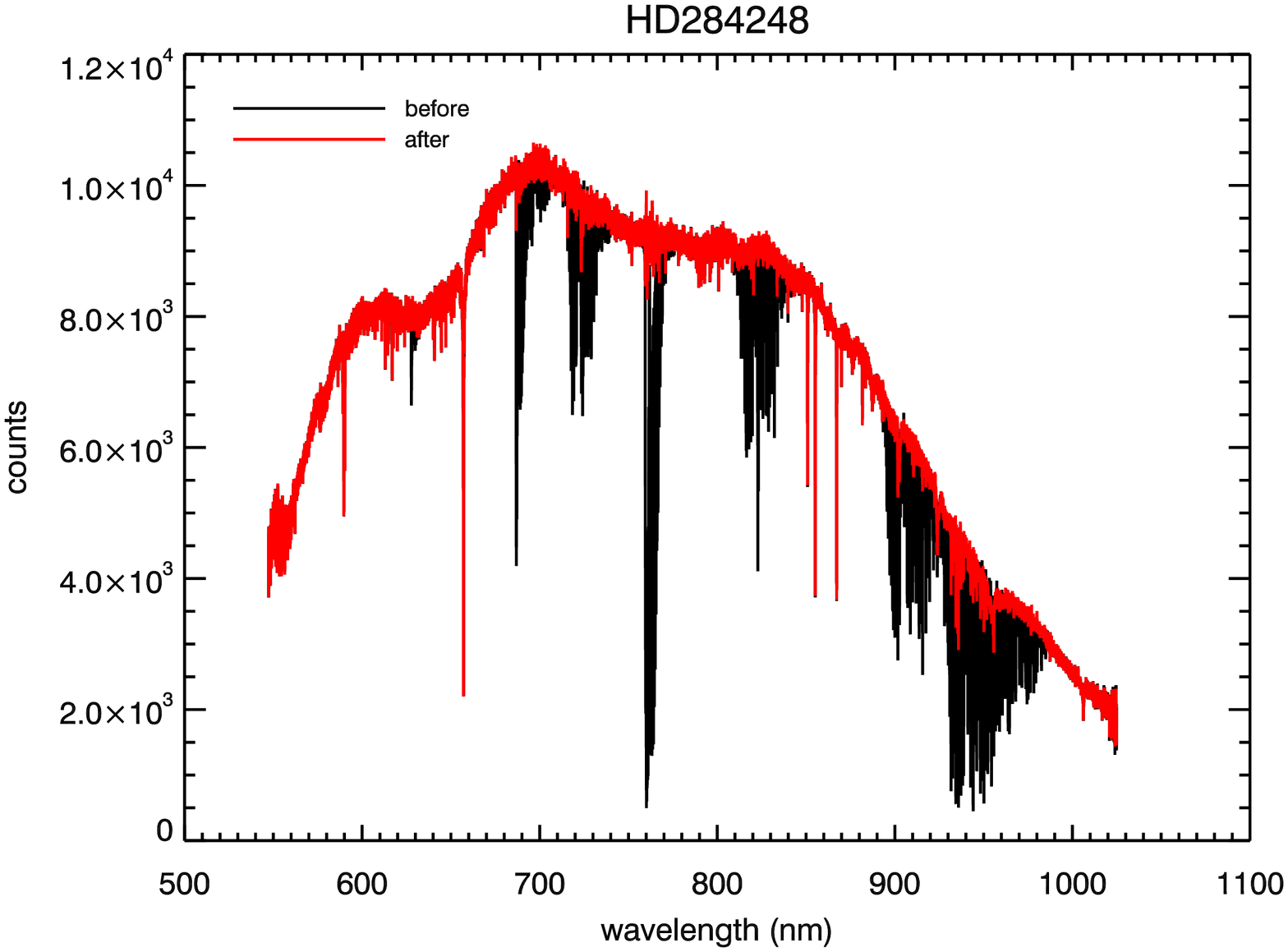}}\hspace*{-0.03\textwidth} \qquad
    \subfloat[]
         {\label{eg1-4}\includegraphics[angle=90,scale=0.25]{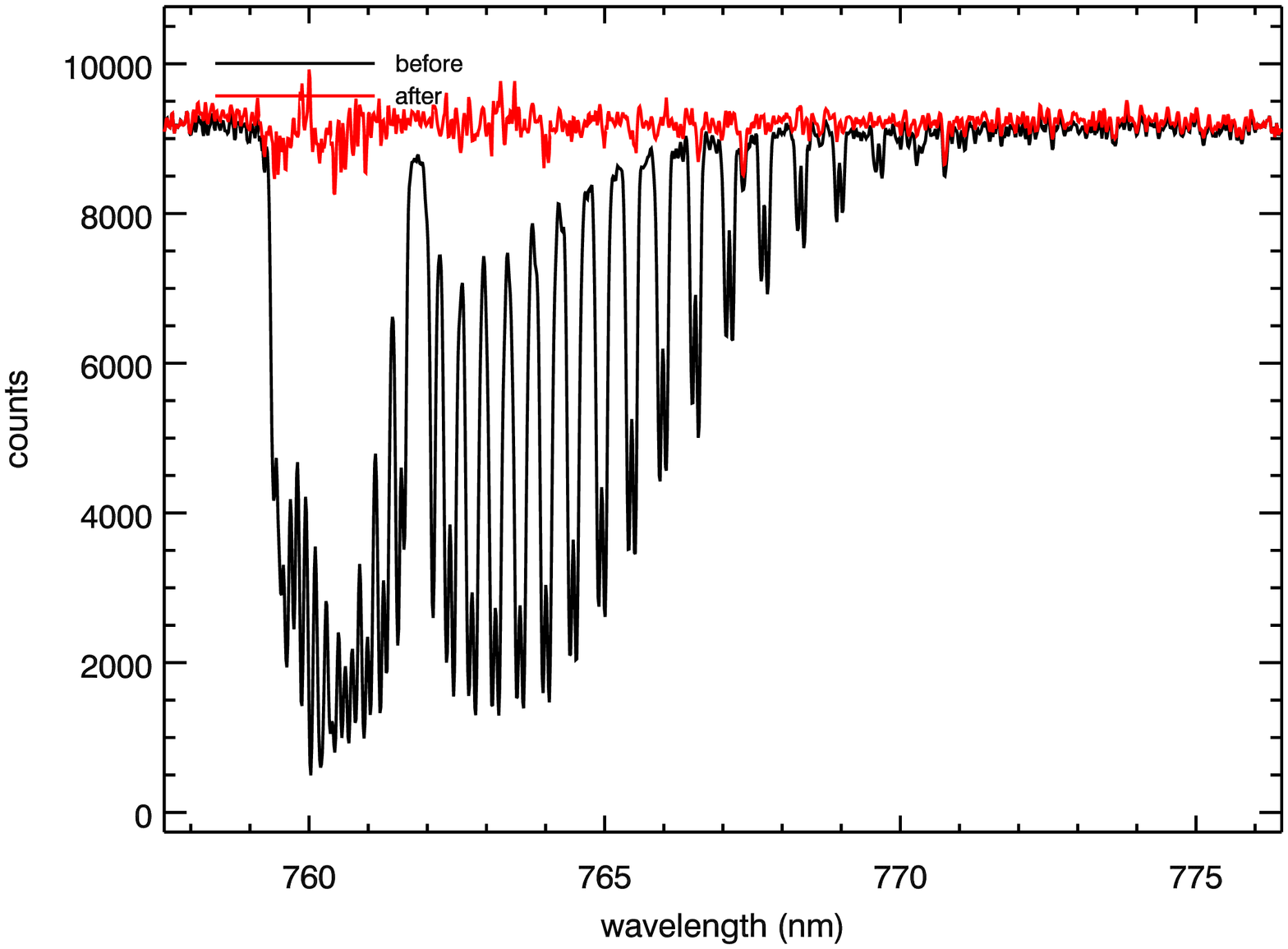}}\hspace*{-0.05\textwidth} \qquad
    \subfloat[]
         {\label{flx}\includegraphics[angle=90,scale=0.25]{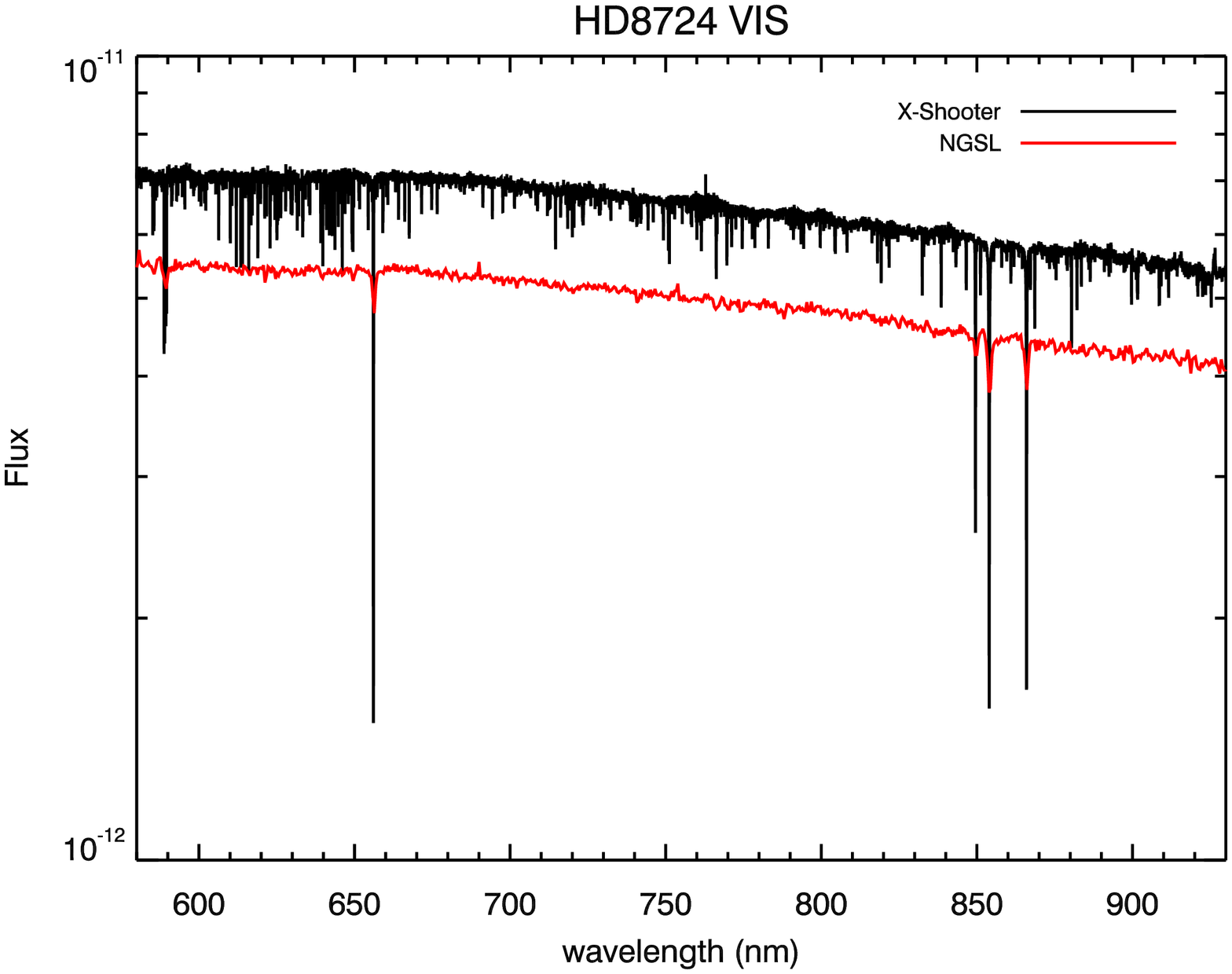}}\qquad
    \caption{Telluric correction and flux calibration of XSL spectra. (a): Telluric corrected version (red) of HD
          284248 together with the original spectrum with different telluric absorption bands (black).
          (b): Same as in Fig. \ref{eg0}, zoomed in wavelength range ${\rm \lambda\lambda 758-776nm}$.
          (c): Flux-calibrated spectrum of HD 8724. The X-Shooter spectrum is shown
           in black and the NGSL spectrum is shown in red. }
        \label{3examples}
\end{figure}



For each science object, we defined several telluric absorption
regions and ran the spectral-fitting program pPXF \cite{ppxf} to find
a ``best fit'' telluric correction spectrum.  We fitted the science
spectrum with a linear combination of the spectra in the (normalized)
telluric library, fitting at the same time the line-broadening of the
science spectra, and multiplying the result with a higher order
polynomial. This step gave a ``best fit'' for the telluric absorption
features to every science object. We then divided the science spectrum
by this ``best fit'', resulting in a telluric-corrected science
spectrum. Fig.~\ref{eg0} shows an example of HD 284248 in the VIS arm
(without flux calibration), where the black line is the original
reduced spectrum before telluric correction and the red line is after
correction. To show the quality of the correction in more detail, we
zoom into the telluric-corrected spectrum in atmospheric ``A'' band
between 758 and 776 nm in Fig.~\ref{eg1-4}.




\subsection{Flux calibration}

To perform a reliable flux calibration, we observed several
spectrophotometric standards (BD+17 4708, GD71, LTT1020, GD153, EG274
and EG21) with a wide ($5.0^{\prime\prime}$) slit in ``stare'' mode at
different airmasses. These flux standard star were then reduced and
extracted through the procedures mentioned above. The spectra were
compared with the flux tables of the appropriate stars from the
CALSPEC HST database
\cite{Bohlin07}\footnote{\url{http://www.stsci.edu/hst/observatory/cdbs/calspec.html}}
and averaged to produce the response function. The Paranal extinction
curve \cite{Patat11} was applied to perform the atmosphere extinction
correction.
Fig.~\ref{flx} shows the flux-calibrated VIS arm spectrum of HD 8724
from X-Shooter (black line) and NGSL (red line). The
UVB arm spectra are corrected in a similar matter.  The NIR arm spectra
will be telluric- and flux-calibrated simultaneously using the hot
star telluric correction spectra in combination with the flux
standards.


\section{Summary}

XSL, the X-Shooter Stellar Library, is intended to be the largest
stellar library with complete wavelength coverage from 320--2480 nm
and covering as large a range in stellar atmospheric parameters as
possible. We will soon have a first version of XSL with spectra of
$\sim240$ stars.  The final library will contain $\sim600$ stars at
moderate resolution ($R\sim10000$). The stellar parameter coverage
will be enforced through carefully selected samples, so that not only
stars with solar metallicities and abundances are included but also
metal-poor or metal-rich stars from the Bulge and Magellanic
Clouds. With X-Shooter's unique capability, the variable stars will be
observed consistently, which will yield more reliable stellar
population models.

\subsection*{Acknowledgments}

We thank our collaborators on the first phase of XSL, D. Silva and
P. Prugniel, and B. Davies and M. Koleva for useful discussions.  We
would also like to extend our great thanks to V. Manieri,
A. Modigliani, J. Vernet, and the ESO staff for their help during the
XSL observations and reduction process.
\section*{References}

\end{document}